\documentclass{article}

\PassOptionsToPackage{numbers}{natbib}

\usepackage[final]{nips_2017}

\usepackage[utf8]{inputenc} 
\usepackage[T1]{fontenc}    
\usepackage{hyperref}       
\usepackage{url}            
\usepackage{booktabs}       
\usepackage{amsfonts}       
\usepackage{nicefrac}       
\usepackage{microtype}      

\usepackage{graphicx}
\usepackage{amsmath, amssymb, bbold}
\usepackage{algorithm}
\usepackage{algorithmic}

\usepackage{subcaption}  
\graphicspath{{./figs/}}

\usepackage{todonotes}

\title{A Goal-Based Movement Model for Continuous Multi-Agent Tasks}

\author{
Shariq Iqbal\\
Duke Institute for Brain Sciences,\\
Duke University\\
Durham, NC 27708\\
\texttt{shariq.iqbal@duke.edu} \\
\And
John Pearson\\
Duke Institute for Brain Sciences,\\
Duke University\\
Durham, NC 27708\\
\texttt{john.pearson@duke.edu} \\
}

\begin{document}

\maketitle

\begin{abstract}
    Despite increasing attention paid to the need for fast, scalable methods to analyze next-generation neuroscience data, comparatively little attention has been paid to the development of similar methods for behavioral analysis. Just as the volume and complexity of brain data have grown, behavioral paradigms in systems neuroscience have likewise become more naturalistic and less constrained, necessitating an increase in the flexibility and scalability of the models used to study them. In particular, key assumptions made in the analysis of typical decision paradigms --- optimality; analytic tractability; discrete, low-dimensional action spaces --- may be untenable in richer tasks. Here, using the case of a two-player, real-time, continuous strategic game as an example, we show how the use of modern machine learning methods allows us to relax each of these assumptions. Following an inverse reinforcement learning approach, we are able to succinctly characterize the joint distribution over players' actions via a generative model that allows us to simulate realistic game play. We compare simulated play from a number of generative time series models and show that ours successfully resists mode collapse while generating trajectories with the rich variability of real behavior. Together, these methods offer a rich class of models for the analysis of continuous action tasks at the single-trial level.
\end{abstract}

\section{Introduction}
As the ability to collect larger volumes of increasingly complex neural data increases, so has the interest of neuroscientists in paradigms that investigate complex, naturalistic behavior \citep{pearson2014decision}. However, typical analysis methods in systems neuroscience often require that experimenters identify in advance specific events around which to collect and average data across trials. Thus, while much attention has been paid to the development of algorithms for flexibly analyzing large-scale brain data (e.g., \citep{freeman2014mapping, pnevmatikakis2016simultaneous, pandarinath2017inferring}), much less effort has been devoted to the analysis of rich behavioral data, particularly multi-agent data. Here, we propose a new method for inferring latent goals from multi-agent behavioral data in the form of multivariate time series. Taking as our example movement data from a dynamic two-player task, we show that these latent goals can furnish a parsimonious account of players' interactions while decoupling intentions from the control needed to execute them.

Most previous work in the decision science and psychology community has used behavioral task designs that either (a) discretize the action and state spaces in order to reduce the dynamics to a discrete choice problem (e.g., \cite{zheng2016generating}) or (b) focus on theoretically-motivated rules that give rise to complex behaviors (e.g., \cite{moussaid2009experimental}). Here, we focus on a different problem, one that edges closer to natural behavior: tasks that involve a continuous state space, controlled by continuous (joystick) inputs, and simultaneous movement by both players. This type of task presents a set of challenges that have received comparatively little treatment in the neuroscience literature: (1) unconstrained behavior that does not easily separate into discrete categories, (2) a lack of meaningful variables against which to align neural data and average over, and (3) complexity that surpasses the capability of simple, interpretable linear models to capture. Our approach strives to solve these problems, using inspiration from the inverse reinforcement learning literature \cite{ng2000algorithms, abbeel2004apprenticeship, Dvijotham2010-lz}, which attempts to use a model to replicate behaviors that match exemplars generated by an expert demonstrator.

We propose a model that learns policies based on the instantaneous goals of each player. These goals are in turn generated based on an underlying joint value function for the two players. In particular, we are interested in abstracting away details of motor planning (the control problem), focusing instead on modeling the evolution of each player's goal, represented as a desired onscreen position. These goals are chosen at each instant based on both the current goal and state of the game, and their values drive joystick input via a simple control model. The resulting value functions are both parsimonious as a description of the game's dynamics and useful as a means of understanding players' choices, since in cases where goals are onscreen locations, they can be directly visualized. Furthermore, goals and value functions provide useful correlates for neural analysis, including trial likelihood, instantaneous expected value (for each player), and entropy.

The outline of the paper is as follows: In Section \ref{data_and_control_section}, we outline our assumptions about the data and control model. In Section \ref{goal_section}, we describe our model for the latent goal states based on Gaussian mixtures. In Section \ref{inference_section} we describe our full generative model for the data along with a variational Bayes algorithm for performing approximate inference on the unknown model parameters.
In Section \ref{results_section} we describe our training procedure and compare the behavior of our model with the dynamics exhibited by real players in the same task.
Section \ref{related} references related work and details our novel contributions. Section \ref{conclusion_section} concludes and discusses potential applications of our model.

\section{Data and control model}
\subsection{Task}
\begin{figure*}[ht]
\begin{center}
    \begin{subfigure}[t]{0.47\textwidth}
        \centerline{\includegraphics[width=\textwidth]{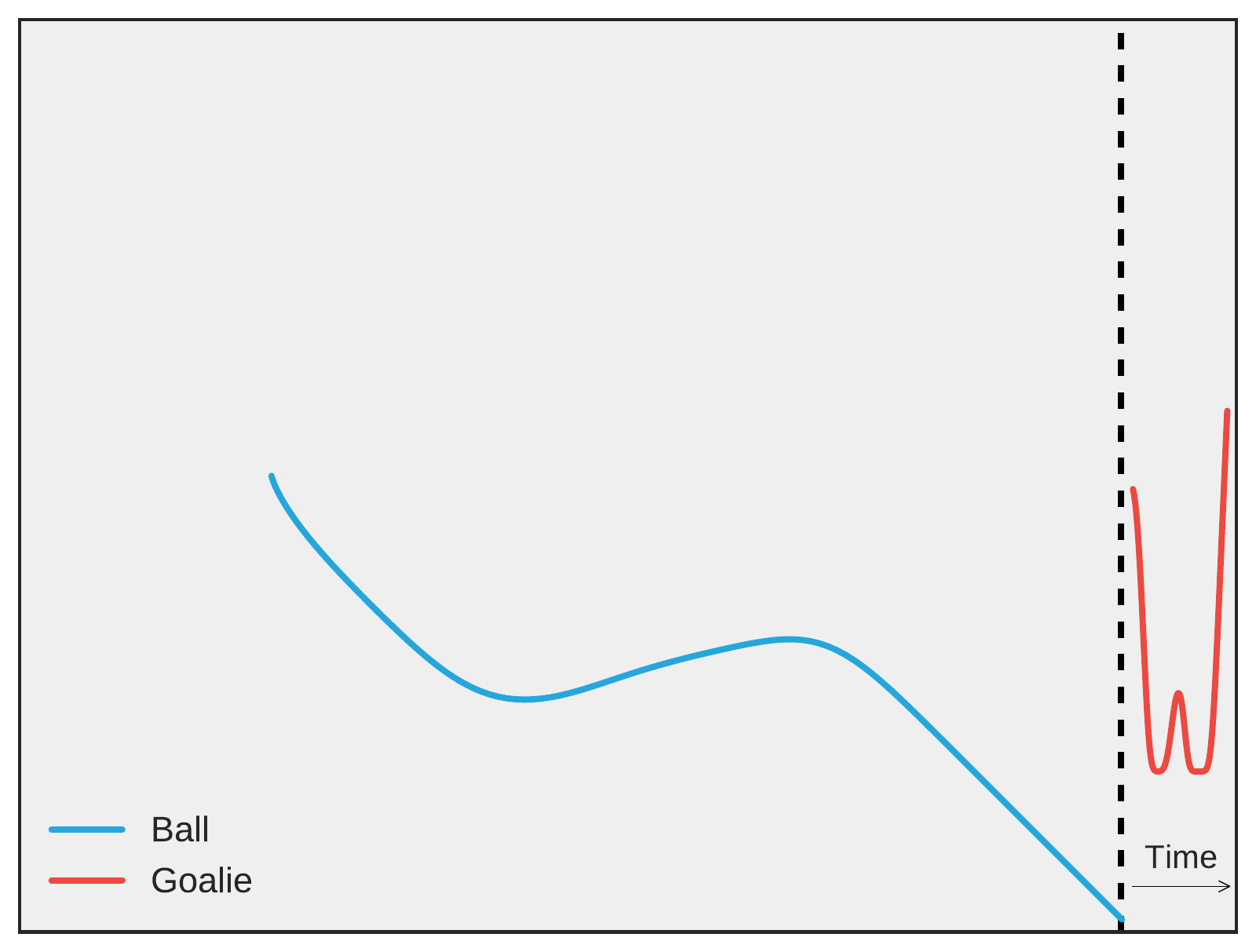}}
        \captionsetup{width=.9\linewidth}
        \caption{Game play. Illustration of onscreen play for the two-player game. The ball (blue) is free to move in two dimensions, while the goalie (red) moves only up or down. Axes are x and y screen directions. Solid lines indicate player trajectories for a single trial. The goalie's trajectory has been stretched along the x axis to indicate its movement through time. For animations of game play, see Supplement.}
        \label{task_fig}
    \end{subfigure}
    \begin{subfigure}[t]{0.47\textwidth}
        \centerline{\includegraphics[width=\textwidth]{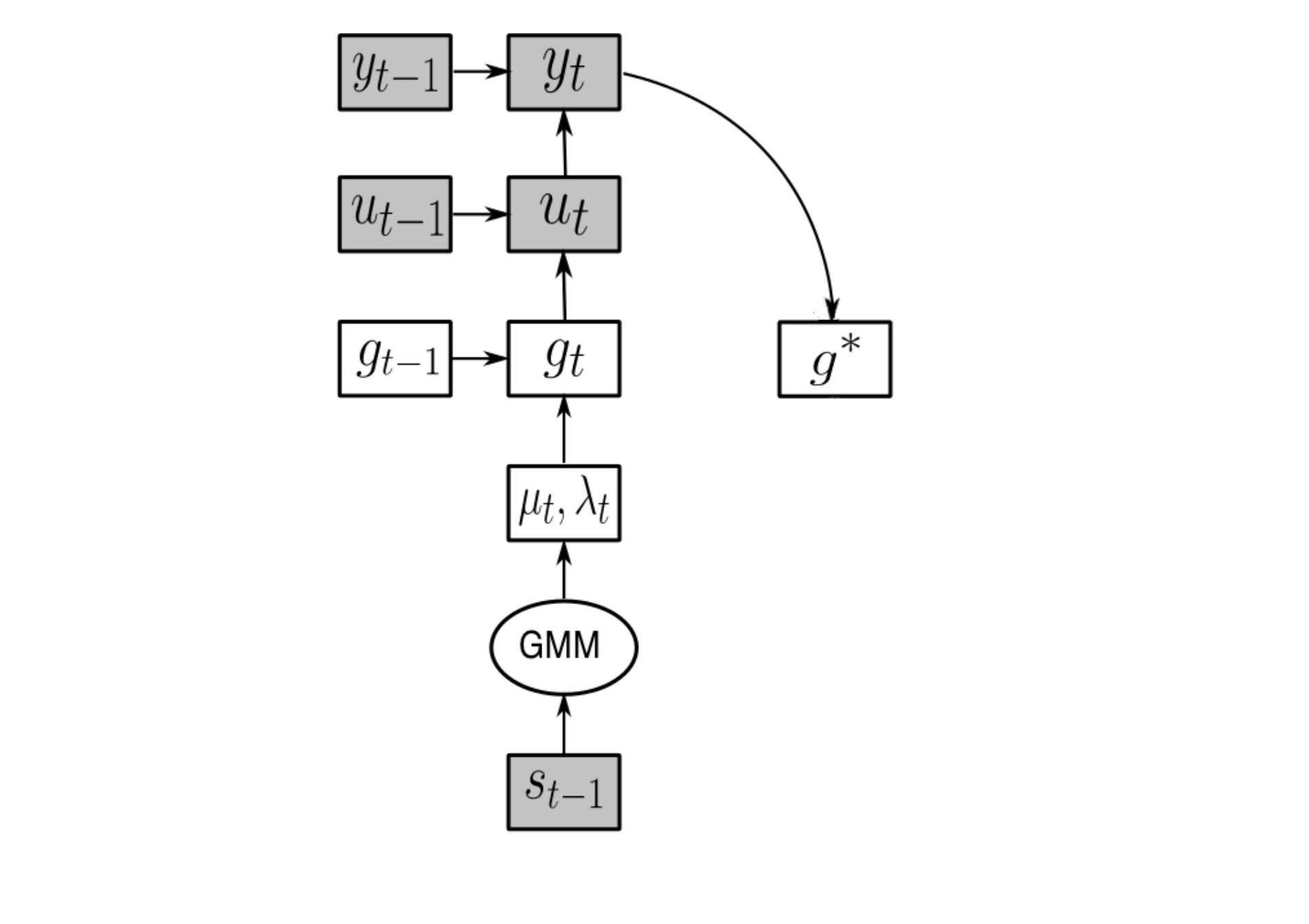}}
        \captionsetup{width=.9\linewidth}
        \caption{Model architecture. Observable trajectories $y_t$ are generated sequentially from control signals $u_t$, which in turn derive from goals $g_t$. In the generative model, parameters of the equations governing goal evolution are drawn from the GMM, which is conditioned on state variables $s_t$. The recognition model takes in $y_t$ and returns posterior samples of the goals.}
        \label{model_fig}
    \end{subfigure}
\end{center}
\vskip -0.1in
\caption{Model architecture and data.}
\end{figure*}

Our data consist of 10 sessions ($\approx 7000$ bouts) of a simple two-player ``penalty shot'' video game played by pairs of rhesus macaques in a neuroscience experiment. Each player used a joystick to control either a ``ball'' able to move in the x and y directions or a ``goalie'' only able to move in the y direction (Figure \ref{task_fig}; \href{https://web.duke.edu/mind/level2/faculty/pearson/assets/videos/penaltyshot/real_trial.mp4}{\color{blue}{movie}}). The objective of the player controlling the puck was to move it across a goal line at the right-hand side of the screen, while the objective of the goalie was to intercept the puck. Each bout of play ended when one of these two outcomes obtained. This game can alternatively be thought of as a modified version of the Atari game ``Pong'', with one player directly controlling the ball and no rebound from the paddle.

\subsection{Model Intuition}
\label{data_and_control_section}
The aim of our model is to use a set of observed movement trajectories from each player in the game to infer underlying intentions. That is, we assume that at each moment, each player has a latent goal, an onscreen position toward which he intends to move his avatar. These goals drive joystick inputs via a simple control model, with the goals of each player co-evolving in time according to some joint dynamics. Control signals are translated by the ``physics'' of the task code into the onscreen movement of the player avatars, which then inform the players' choices of goals in at the next time step. The result is a generative model capable of producing entirely new behavior that captures the variability present in real opponents. For our case study, the data consist of onscreen trajectories of player-controlled avatars, concatenated into a multivariate time series, but we emphasize that our model is more generally applicable to any multivariate time series in which the assumption of an underlying time- and state-dependent set-point for a control process is valid.

\subsection{Data}
Data consisted of onscreen positions of both avatars at regularly sampled time points $y_t = (x_{puck}, y_{puck}, y_{goalie})$ for each bout of play (Figure \ref{task_fig}). We normalize each coordinate such that $(y_t)_i \in [-1, 1]$, and we also consider a state variable for the system $s_t$ on which agents' behavior might be conditioned. For this work, we restrict this information to consist of instantaneous positions and velocities for both players: $s_t \equiv (y_t, y_t - y_{t-1})$. We assume that the training set consists of a large number of bouts of play, with each bout constituting a single realization of each time series.\footnote{In this work, we assume these realizations are iid, though future work might consider modeling changes in the value function of each agent during the course of play.} We further assume that control inputs are translated into onscreen variables via
\begin{equation}
    \label{eom}
    y_{t} = y_{t-1} + v_\mathrm{max} \odot \sigma(u_t)
\end{equation}
with $v_\mathrm{max}$ a scaling between control and velocity, $\sigma(x)$ a sigmoid function reflecting saturation of control signals, and $\odot$ elementwise multiplication.

\subsection{Control model}
We assume that at each moment in time, each agent has a desired goal state $g_t$ (here an onscreen position) and a controller capable of achieving this state by minimizing the error $e_t \equiv g_t - y_{t-1}$. For simplicity, we assume that this minimization is performed by a proportional integral derivative (PID) controller, which takes the discrete form (for a single control variable)
\begin{equation}
    \label{PID}
    \Delta u_t = K_p
    \left[
        \left(1 + \frac{\Delta t}{T_i} + \frac{T_d}{\Delta t}\right) e_t -
        \left(-1 - \frac{2T_d}{\Delta t}\right) e_{t - 1} +
        \frac{T_d}{\Delta t} e_{t - 2}
    \right]
\end{equation}
where $K_p$ is the proportional control constant, $\Delta t$ is the time step, and $T_i$ and $T_d$ are the integration and derivative time scales, respectively. More generally, we can write the change in control signal as a convolution: $\Delta u = L * (g - y)$, with $L$ given by the filter defined by $(K_p, T_d, T_i)$ in (\ref{PID}). In what follows, we will also assume some Gaussian uncertainty in the control signal
\begin{equation}
    \label{umodel}
    u_{t} \sim \mathcal{N}(u_{t-1} + L * (g_t - y_{t-1}), \epsilon^2)
\end{equation}
while keeping the relationship (\ref{eom}) between $u_t$ and $y_t$ deterministic. Here, we have written the update for a single variable. In the multivariate case, we assume control (and control noise) is independent for each dimension of the combined vector $\mathbf{u}_t$.

\section{Goal model}
\label{goal_section}
For the goal time series, we will assume a Markov process in which new goals are probabilistically selected at each time based on both the current goal and the current state of the system. That is,
\begin{equation}
    \label{pmodel}
    p(u, g) \propto \prod_t p(u_t|u_{t-1},g_t, y_{t-1})\, p(g_t|g_{t - 1}, s_{t - 1})
\end{equation}
More specifically, we will assume that at each time point, there exists a function $V(g_{t}, s_{t-1})$ that captures the benefit in setting a particular goal at the next time step based on the current state of the system. That is, we want to increase $V$ as often as possible. However, we add as a regularization constraint the idea that there should be some cost to large changes in goals, which we take to be quadratic in the distance between successive points. Explicitly, let
\begin{equation}
    \label{joint_model}
    \log p(u, g) = \sum_t \left[
        -\frac{1}{2\epsilon^2} \lVert u_t - \bar{u}_{t}(g_t, y_{t-1}) \rVert^2\right.
        \left. -\frac{1}{2\sigma^2} \lVert g_t - g_{t - 1} \rVert^2
        + V(g_t, s_{t - 1})
    \right] - \log Z
\end{equation}
Here, $\bar{u}_{t} \equiv u_{t-1} + L * (g_t - y_{t -1})$ is the the predicted control and $\epsilon$ and $\sigma$ govern the noise in the observations and goal diffusion, respectively. In what follows, we will also find it useful to define $\beta \equiv \sigma^{-2}$ in order to write $\log p(g) = -\beta E(g|s) - \log Z$ with
\begin{equation}
    \label{energy_model}
    E(g|s) = \sum_t \left[\frac{1}{2}\lVert g_t - g_{t-1} \rVert^2 - \sigma^2 V(g_t, s_{t-1})\right]
\end{equation}

This formulation admits multiple interpretations: In analogy with the path integral formulation of stochastic processes, it can be viewed as a model in which the ``potential energy'' $U(g) \equiv -\sigma^2 V$ trades off at each time step with the ``kinetic energy'' $K \equiv \dot{g}^2/2$. In the limit of small $V$/small $\sigma$/large $\beta$, one is in either the low-temperature thermodynamic limit or the high-mass classical limit, and $g$ is a spatially-varying Gaussian process. Alternately, in the limit of large $\sigma$, goals are simply chosen independently at each time point. In any case, we have made the strong assumption that dependence of $g_t$ on $g_{t-1}$ occurs only through a momentum term, requiring that the ``static'' $V$ term carries most of the weight of explanation.

Unfortunately, for general $V(g|s)$, the distribution implied by (\ref{energy_model}) is of the Boltzmann-Gibbs form and is impossible to sample efficiently. And while it is possible in principle to train a generative network to sample innovations in the $g$ time series directly, the goal of our inference is to model $V$ itself, since this captures the strategic interplay between the two opponents' goals.

\subsection{$V(g)$ as a Gaussian Mixture Model}
However, for any $V$ defining an absolutely continuous Boltzmann distribution it is possible to approximate the potential energy piece of (\ref{energy_model}) as a finite mixture of Gaussians \citep{park1991universal}:
\begin{equation}
    \label{mixture_model}
    e^{V(g)} = \sum_{k = 1}^K w_k e^{-\frac{\beta }{2} (g - \mu_k)^\top \Lambda_k (g - \mu_k)}\sqrt{\frac{\beta\Lambda_k}{2\pi}}
\end{equation}
with $\sum_k w_k = 1$. Now, sampling from $e^V$ is simply a matter of first sampling a mode $k$ from the mixture and subsequently taking $g_t$ from
\begin{equation}
    \label{pg_given}
    p(g_t|g_{t-1}, \mu_t, \Lambda_t, \sigma^2, k) \propto
    \exp \left(
        -\frac{1}{2\sigma^2}\left[
        \lVert g_t - g_{t-1} \rVert^2
        + (g_t - \mu_{kt})^\top \Lambda_{kt} (g_t - \mu_{kt})
    \right]
    \right)
\end{equation}
which is equivalent to
\begin{equation}
    \label{next_g_draw}
    g_{t}|g_{t-1}, \Lambda_t, \mu_t, \sigma^2, k \sim \mathcal{N}\left((\mathbb{1} + \Lambda_k(s))^{-1}(g_{-1} + \Lambda_k(s)\mu_k(s)), \sigma^2 (\mathbb{1} + \Lambda_k)^{-1} \right)
\end{equation}
That is, conditioned on previous goal position, $\mu$, $\Lambda$, and $k$, the new goal is simply governed by a normal distribution. This in turn implies that the distribution of $g_t$ with $k$ marginalized out is itself a mixture of normals with the same weights $w_k$. Thus, given that we can efficiently sample $(\mu_t, \Lambda_t)$, we have a means of sampling entire goal trajectories.

We model each of the parameters $\mu$, $\Lambda$, and $w$ of our Gaussian Mixture Model with a multi-layer perceptron that takes the state of the system as its input and outputs a parameter for each mode $k$. This allows for our model to learn complex nonlinear relationships between state and value while remaining easy to sample from. Though we have defined the model with a full covariance structure, for the results below, we used a diagonal covariance $\Lambda_k = \mathrm{diag}(\boldsymbol{\lambda}_k)$. Likewise, we use a Gaussian mixture for the initial distribution over goals: $g_0 \sim \mathrm{GMM}(\mu_0, \Lambda_0, w_0)$.

\section{Full model and inference}
\label{inference_section}
\subsection{Full generative model}
Putting together the assumptions of the previous sections, we can now present a full generative model for our data (Figure \ref{model_fig}). A given trial can be generated timepoint-by-timepoint according to Algorithm \ref{alg_generate}.
\begin{algorithm}[ht]
\caption{Generative model for trajectory data}\label{alg_generate}
\begin{algorithmic}
    \REQUIRE $\sigma$, $\epsilon$, $L$, $v_\mathrm{max}$
    \STATE Initialize $y_0, s_0, u_0$
    \STATE $g_0 \sim G_{g_0}$ \COMMENT{Sample from distribution over initial goals}
    \FOR{$t = 1 \ldots T$}
    \STATE $k \sim \mathrm{Categorical(\mathbf{w})}$ \COMMENT{Sample mode of GMM}
    \STATE $(\mu_t, \Lambda_t) = (\mu_k(s_{t - 1}), \Lambda_k(s_{t-1}))$
    \STATE $g_{t} \sim \mathcal{N}\left((\mathbb{1} + \Lambda_t)^{-1}(g_{t-1} + \Lambda_t\mu_t), \sigma^2 (\mathbb{1} + \Lambda_t)^{-1} \right)$ \COMMENT{Sample new goal}
    \STATE $e_t \gets g_t - y_{t-1}$ \COMMENT{Control error}
    \STATE $u_t \sim \mathcal{N}(u_{t-1} + L * e_t, \epsilon^2)$ \COMMENT{PID control}
    \STATE $y_t \gets y_{t-1} + v_\mathrm{max} \tanh(u_t)$ \COMMENT{Update position}
    \STATE $s_t \gets (y_t, y_t - y_{t -1}, \ldots)$ \COMMENT{Update state}
    \ENDFOR
\end{algorithmic}
\end{algorithm}

\subsection{Approximate Bayesian inference}
\label{bayes_inference_section}
Given the observed system trajectory $y_t$ (equivalently, the observed control signal $u_t$), we need to make inferences about the underlying goal trajectory $g_t$, and furthermore, the parameters of the value function $\mu$ and $\Lambda$. In general, full Bayesian inference is intractable, but we employ a variational Bayes (VB) approach \cite{beal2003variational, wainwright2008graphical} that approximates this procedure. In brief, VB attempts to minimize the Kullback-Leibler divergence between a known generative model for which inference is intractable, $p(\mathcal{D}, z)$ with data $\mathcal{D}$ and latent variables $z$, and an approximating family of posterior distributions, $q(z)$. This is equivalent to maximizing an evidence lower bound (ELBO) given by
\begin{equation}
    \label{ELBO}
    \mathcal{L} = \mathbb{E}_{q(z)}[\log p(\mathcal{D}, z)] - \mathcal{H}[q(z)] \le \log p(\mathcal{D})
\end{equation}
with $\mathcal{H}[q(z)]$ the entropy of the approximating posterior. That is, inference is transformed into an optimization problem in the parameters of the approximate posterior $q(z)$, amenable to solution by gradient ascent. In our model, we make use of recently developed ``black box'' methods \cite{ranganath2014black, kucukelbir2015automatic, rezende2014stochastic, Kingma2013-td} in which the gradients of the ELBO are replaced with stochastic approximations derived by sampling from $q(z)$, avoiding often difficult computations of the expectation in (\ref{ELBO}). Thus our only requirement for $q(z)$ is that it be straightforward to sample from.

In our case, we begin with the generative model specified by (\ref{pmodel}) and detailed in Algorithm \ref{alg_generate}. For posteriors, we used the variational latent dynamical system (VLDS) from \cite{Archer2015-ec, Gao2016-ck} for $q(g|y)$. As in \cite{rezende2014stochastic, Kingma2013-td} we use samples from the posterior model in order to update \emph{both} the parameters of the generative model ($L$, $\epsilon$, $\sigma$, and the weights of the neural networks parameterizing $\mu(s)$, $\Lambda(s)$, and $w(s)$) and the parameters of the approximate posterior  $q_g$ via gradient ascent.\footnote{We do not explicitly model $q(u_t|g, y)$, though $p(u_t|g, y)$ is normal at each time step. Since $\epsilon \ll 1$, this is effectively a delta function.}.

\section{Case study: Penalty Shot Game}
\label{results_section}

\begin{figure*}[ht]
\begin{center}
    \begin{subfigure}[t]{0.47\textwidth}
        \centerline{\includegraphics[width=\textwidth]{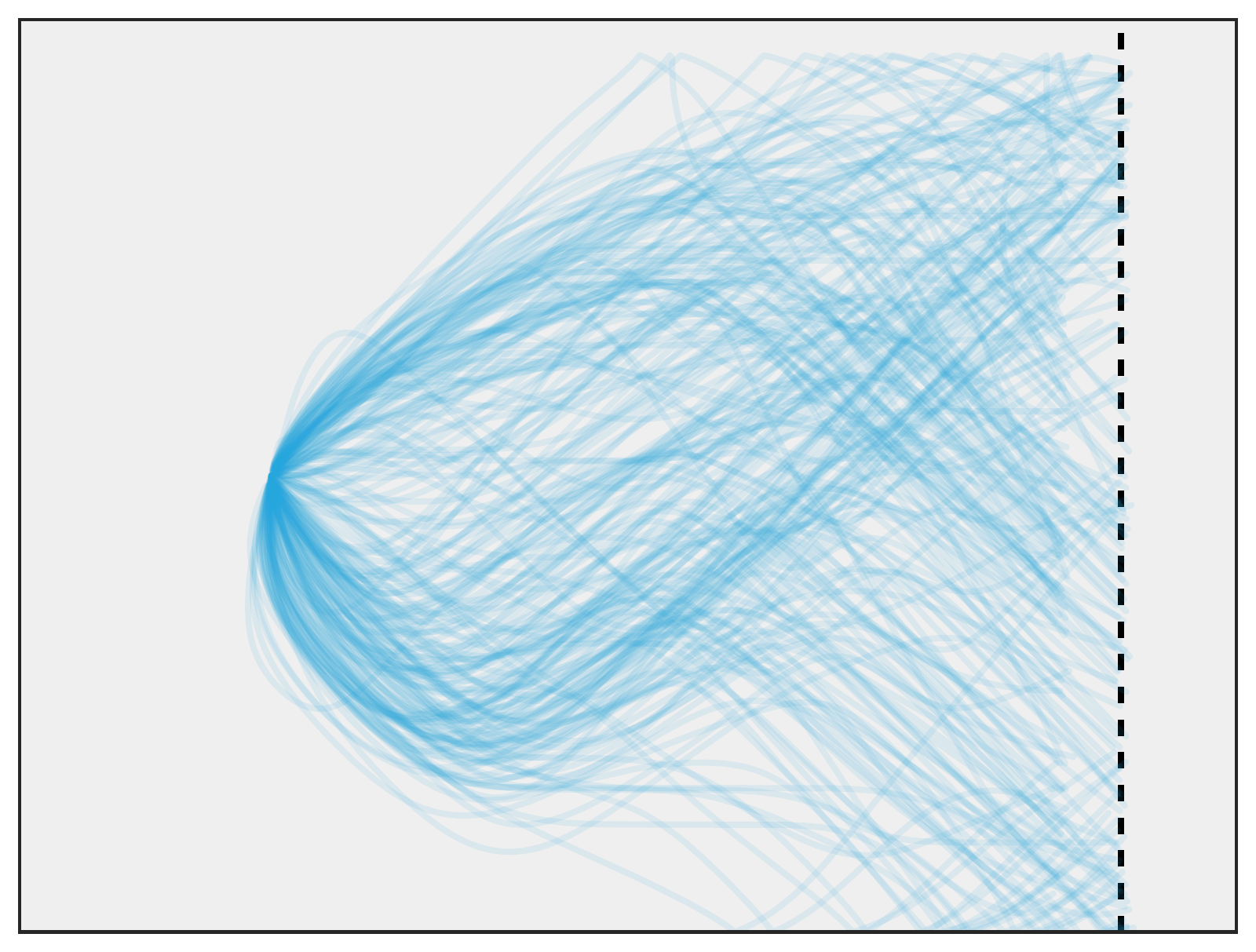}}
        \captionsetup{width=.9\linewidth}
        \caption{Training data}
        \label{real_trajectory_fig}
    \end{subfigure}
    \begin{subfigure}[t]{0.47\textwidth}
        \centerline{\includegraphics[width=\textwidth]{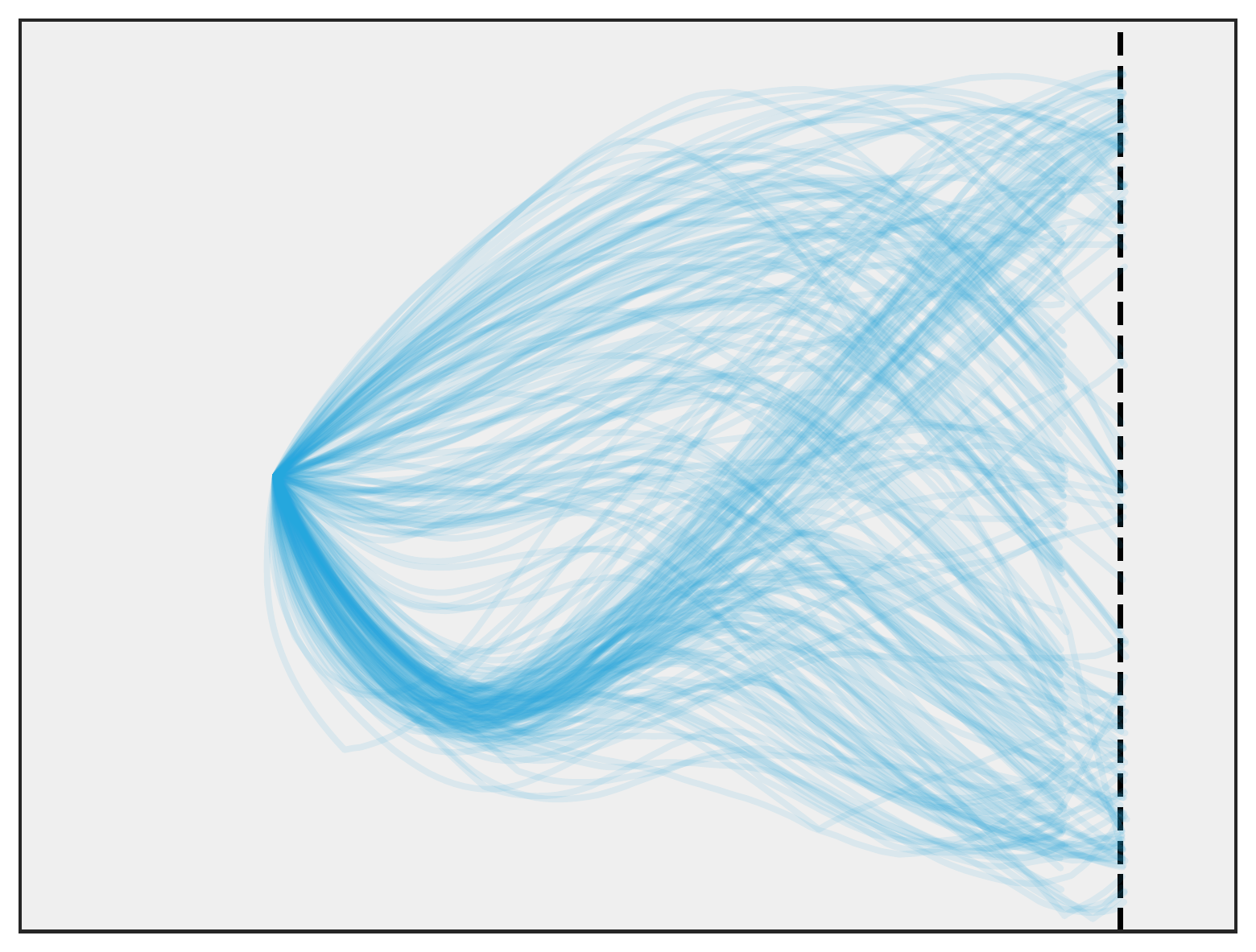}}
        \captionsetup{width=.9\linewidth}
        \caption{Algorithm \ref{alg_generate}}
        \label{gen_trajectory_fig}
    \end{subfigure}
    \begin{subfigure}[b]{0.47\textwidth}
        \centerline{\includegraphics[width=\textwidth]{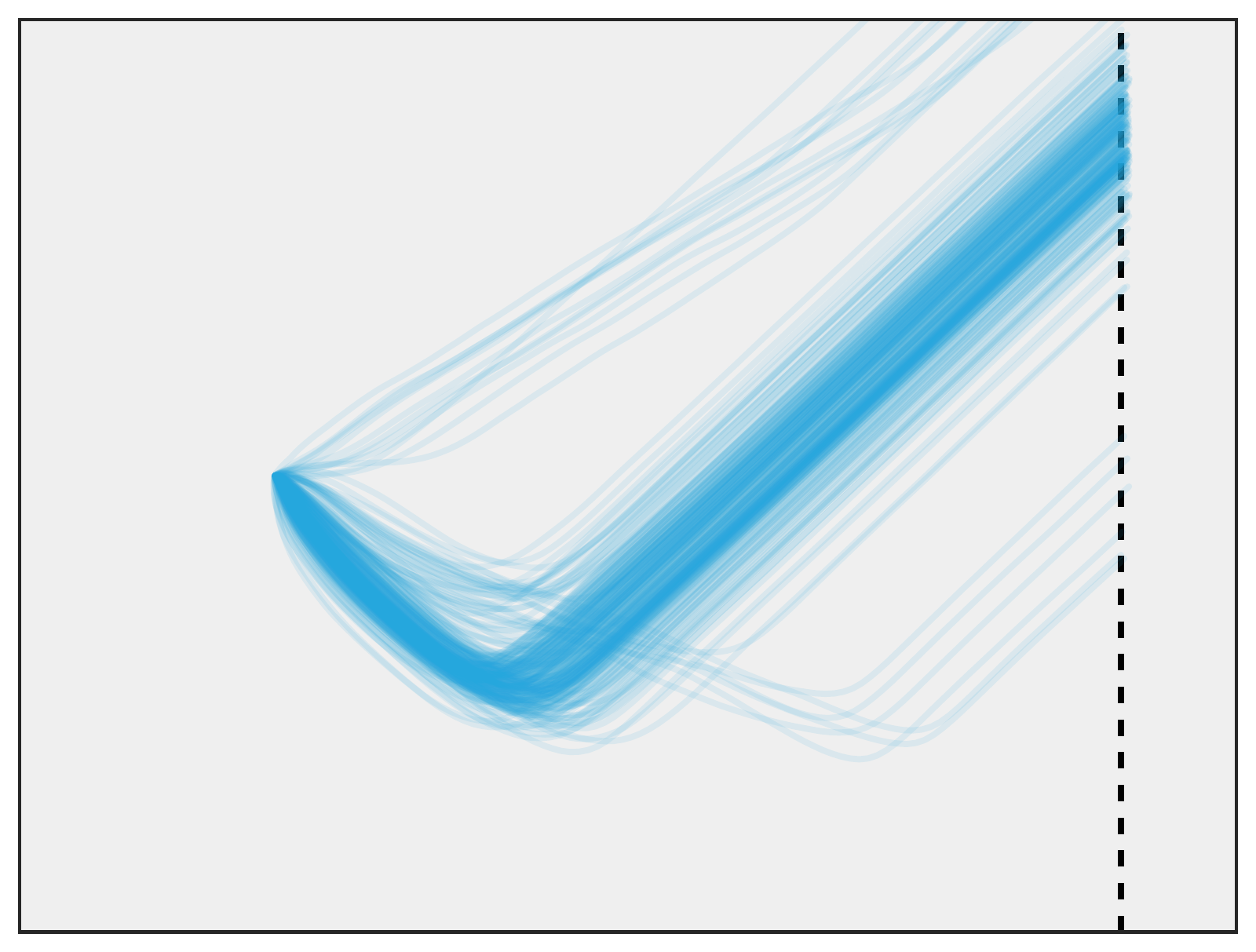}}
        \captionsetup{width=.9\linewidth}
        \caption{Deep Latent Gaussian Model}
        \label{gen_trajectory_vae}
    \end{subfigure}
    \begin{subfigure}[b]{0.47\textwidth}
        \centerline{\includegraphics[width=\textwidth]{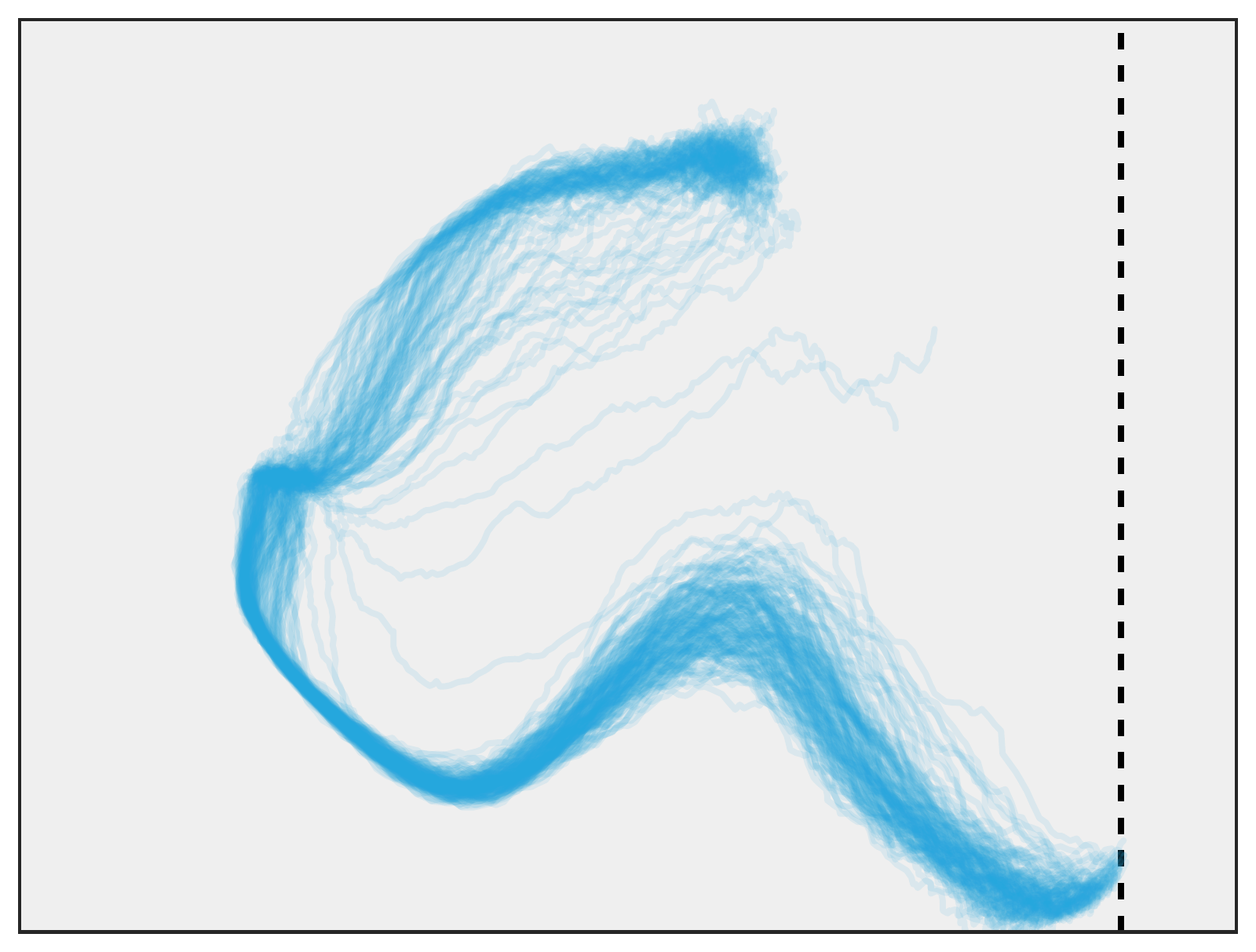}}
        \captionsetup{width=.9\linewidth}
        \caption{Variational RNN}
        \label{gen_trajectory_vrnn}
    \end{subfigure}
\end{center}
\vskip -0.1in
\caption{Comparison of real and generated ball trajectories. 250 samples each from data and generative models. The VAE and VRNN models suffer from mode collapse, while our GMM-based model correctly captures all trajectory types.}
\label{trajectory_fig}
\end{figure*}

\subsection{Training}
Training was done in an end-to-end fashing using variational inference. We trained using a single-trial black box approximation to the evidence lower bound \cite{ranganath2014black, kucukelbir2015automatic}. This consisted of 1) drawing a random trial, 2) sampling inferred goals from $q(g|y)$, 3) using that sample to calculate the evidence lower bound (ELBO), and 4) jointly updating the parameters of our generative and posterior models using gradient ascent with ADAM \cite{Kingma2014-pz} with a learning rate of $10^{-3}$. The model was trained until a smoothed version of the ELBO reached convergence.

For the posterior on $g$ we used a three-layer neural network with 25 hidden units for each component of the mean and a network of the same structure for each nonzero entry of the Cholesky factor of the covariance as in \cite{Archer2015-ec}. We achieved best performance by setting $\sigma \sim 10^{-3}$ and including a regularization penalty for $\epsilon$. Finally, to deal with nonidentifiability in the goal states arising from the tanh function applied to control, we incorporated a hinge loss on goals located outside the visible screen.

For comparison, we also fit our data using a Deep Latent Gaussian Model (DLGM) \citep{rezende2014stochastic} as a model of $V(g)$ and a Variational Recurrent Neural Network (VRNN) \citep{Chung2015-sy} as a model of the empirical time series $y_t$. Our DLGM was also trained jointly with the control model using a Variational Bayes procedure. The generative model used 3 layers, with 25 hidden units, while the posterior model for the mean and variance of each generative layer had 2 layers and 25 hidden units. Rectified linear units were used as the nonlinearity for all hidden layers. The VRNN used \href{https://github.com/jych/nips2015_vrnn}{\color{blue}{code}} provided by the authors, which was modified to work for our data. These modifications included reducing the number of latent variables and hidden units, as well as allowing the model to take in extra inputs that it doesn't need to predict (i.e. we used velocity as an additional input, even though we only care about predicting position). Here, we fit the trajectories ($y_t$) directly, as opposed to the DLGM, which followed the control-goal ($u_t$, $g_t$) formulation of the GMM.

\subsection{Results}
Our model is successfully able to capture the wide variability of trajectories exhibited by player data (Figure \ref{trajectory_fig}). The resemblance of our model's generated data to real data, combined with the structured interpretibility of the model, suggests that our assumptions are at least sufficient as an ``as if'' explanation of dynamics. Moreover, the generative nature of the model allows efficient calculation or approximation of quantities like momentary surprise or entropy of the goal mixture that are of interest as possible neural correlates. For comparison, we also present similar numbers of trials generated from the Deep Latent Gaussian Model (DLGM) \cite{rezende2014stochastic}, a form of probabilistic variational autoencoder, and the Variational Recurrent Neural Network (VRNN) of \cite{Chung2015-sy}. Clearly, both the DLGM (Figure \ref{gen_trajectory_vae}) and the VRNN (Figure \ref{gen_trajectory_vrnn}) suffer from mode collapse, while our GMM-based model (Figure \ref{gen_trajectory_fig}) reproduces all the major trajectory motifs of the real data (Figure \ref{real_trajectory_fig}), while maintaining an interpretable structure. We ascribe the failure of the DLGM and VRNN in capturing the variability in our data to the strong multimodality of trajectories at strategic decision points in the trial. In our experiments, despite initially producing a variety of trajectories both models eventually succumbed to the mode collapse exhibited in (Figures \ref{gen_trajectory_vae} and \ref{gen_trajectory_vrnn}).

\begin{figure*}[!ht]
\vskip 0.2in
\begin{center}
    \begin{subfigure}[t]{0.45\textwidth}
        \centerline{\includegraphics[width=\textwidth]{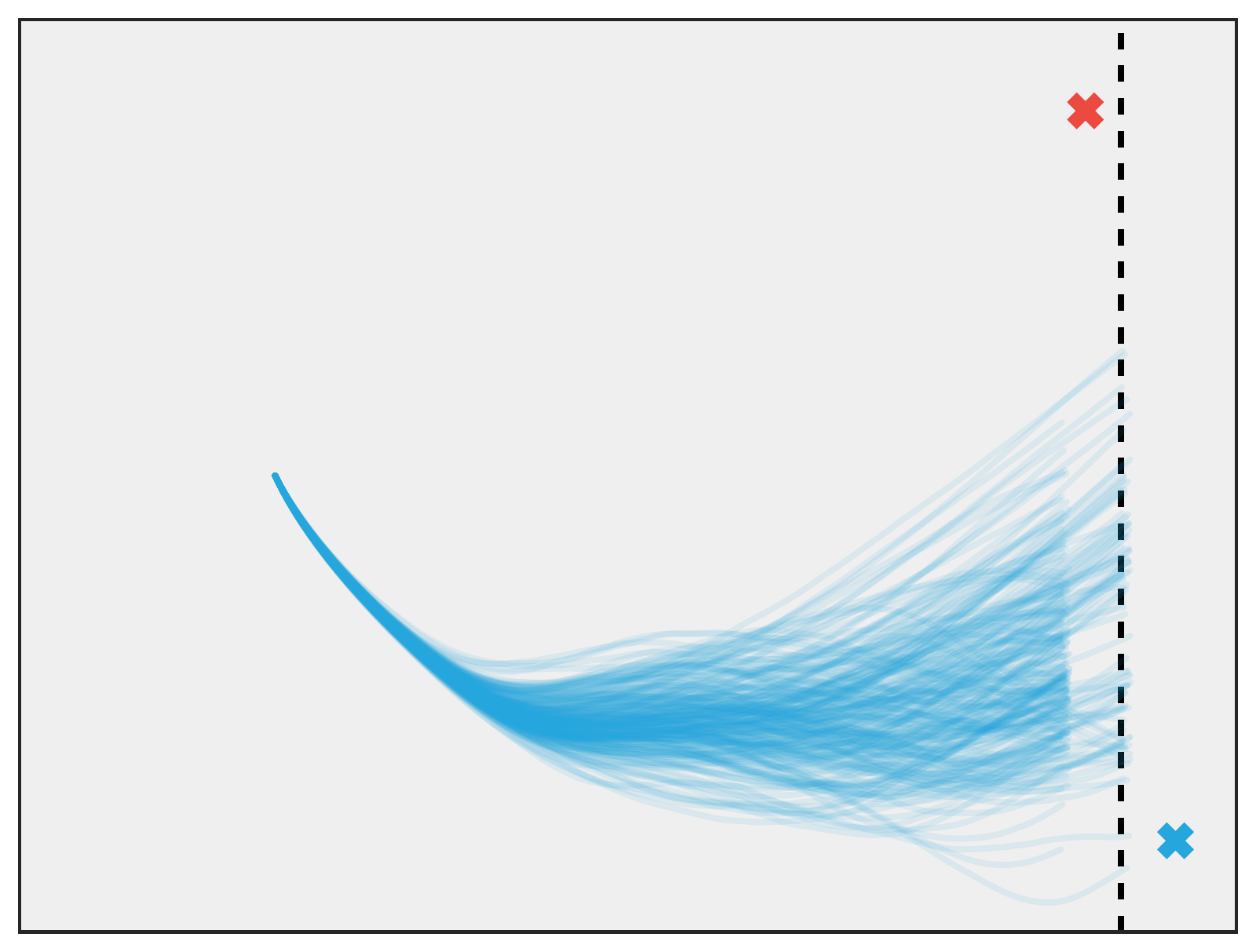}}
        \captionsetup{width=.9\linewidth}
        \caption{Generated ball trajectories with initial goal states for ball in the bottom right corner and the goalie on the top of the screen. Note that the ball generally reaches a decision point late in trial in response to the goalie's downward motion in response to the ball's initial goal.}
        \label{init_goalieup_balldown_fig}
    \end{subfigure}
    \begin{subfigure}[t]{0.45\textwidth}
        \centerline{\includegraphics[width=\textwidth]{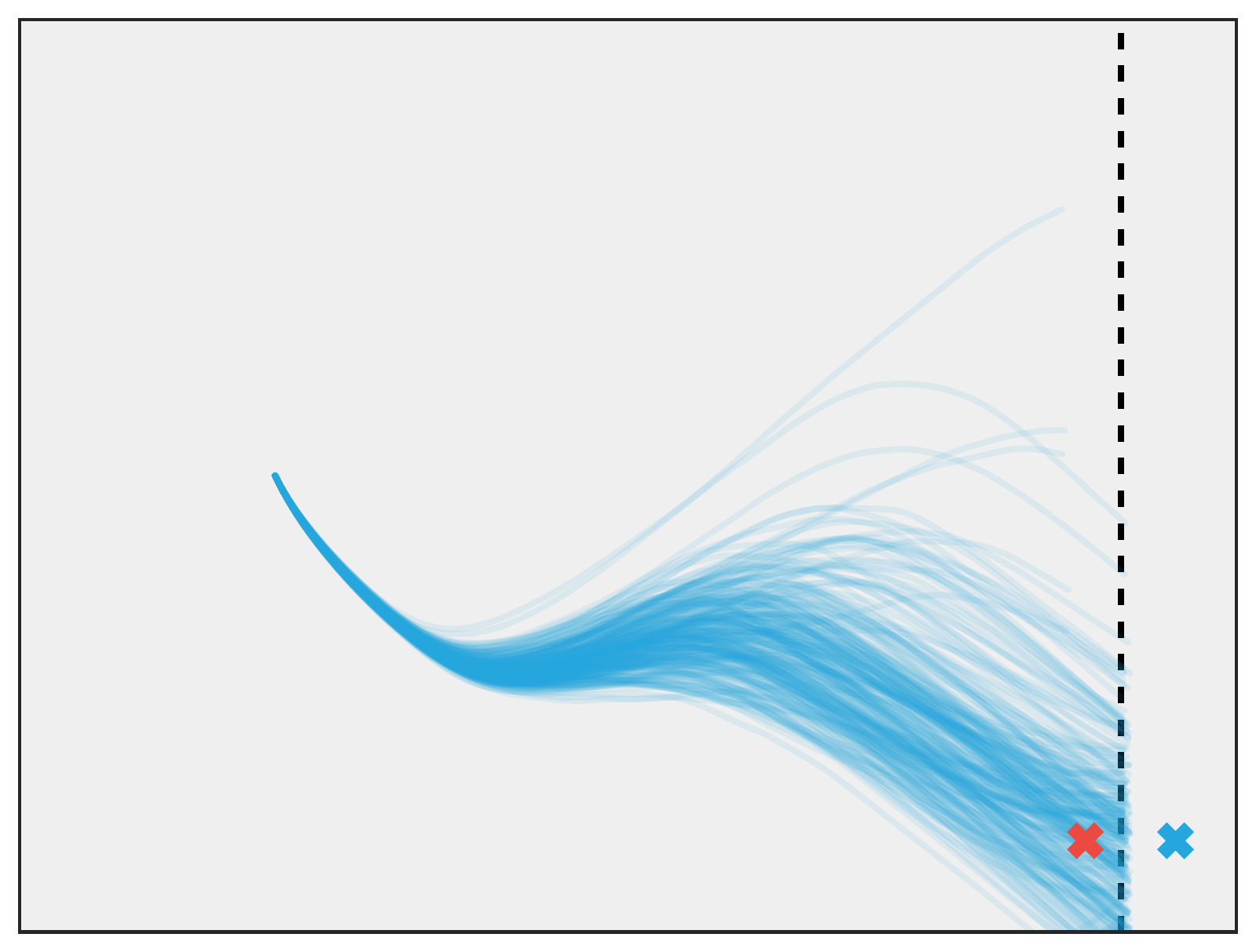}}
        \captionsetup{width=.9\linewidth}
        \caption{Generated ball trajectories with initial goal states for ball in the bottom right corner and the goalie also on the bottom. Here, the ball generally shifts goal earlier because the goalie is initially headed downward, creating an earlier decision point.}
        \label{init_both_down_fig}
    \end{subfigure}
\end{center}
\vskip -0.1in
\caption{Effects of initial goal state.}
\label{init_goal_fig}
\end{figure*}

Based on the raw data (Figure \ref{real_trajectory_fig}), it is clear that the players in our task had very distinct goals at the beginning of each trial. In order to demonstrate the effect of initial goal state on trial outcome, as well as gain insight into the interaction between agents in our model, we generated trials with specified initial goal states. These can be seen in Figure \ref{init_goal_fig}. Red Xs indicate the goalie's initial goal, and blue Xs indicate the ball's initial goal. Towards the beginning of the trial, the agents both strongly follow their initial goal, but as the trial progresses, the opponents' trajectories begin to influence one another. In the case of \ref{init_goalieup_balldown_fig}, the goalie (not shown), having initially guessed incorrectly, quickly turns around. At this point, the ball reaches a decision point, with some subsequent trajectories bending upward and some going down. In the case of \ref{init_both_down_fig}, however, the ball diverges from its initial goal more quickly because the goalie guesses correctly. At this point, the ball either tries to feint up and go down, or, less commonly, sharply turn up. These types of behaviors demonstrate the ability of our model to accurately capture the interaction inherent in the task, as opposed to simply memorizing trajectories.

\subsection{Animations}

In order to visualize the dynamics of our model in real time, we generated animations of individual trials (both real and generated from our model) with different visualizations.

\begin{itemize}
\item {\bf Animation 1:} A real trial, played by monkeys. \href{https://web.duke.edu/mind/level2/faculty/pearson/assets/videos/penaltyshot/real_trial.mp4}{(\color{blue}{link})}
\item {\bf Animation 2:} The same real trial, with goal states inferred by our recognition model at each time point plotted with empty colored shapes corresponding to their agent. \href{https://web.duke.edu/mind/level2/faculty/pearson/assets/videos/penaltyshot/real_trial_goals.mp4}{(\color{blue}{link})}
\item {\bf Animation 3:} A trial generated from our model. \href{https://web.duke.edu/mind/level2/faculty/pearson/assets/videos/penaltyshot/gen_trial.mp4}{(\color{blue}{link})}
\item {\bf Animation 4:} The same generated trial, with corresponding goal states from our generative model. \href{https://web.duke.edu/mind/level2/faculty/pearson/assets/videos/penaltyshot/gen_trial_goals.mp4}{(\color{blue}{link})}
\end{itemize}

\section{Related Work}
\label{related}
There is a long history of investigating behavior in neuroscience from the viewpoint of control models, particularly for eye movements \citep{carpenter1988movements} and motor coordination \citep{todorov2002optimal}. Models based on recurrent neural networks \citep{sussillo2009generating, song2016training} have been successful at generating patterns of activity described as multivariate time series, as have, more recently, spiking models \citep{thalmeier2016learning, abbott2016building}. However, the focus in the first case has been on the control process itself (much simplified in our formulation) and only secondarily on intentions, while in the second class of models, the focus has been on replicating output from a tutor signal rather than reproducing highly irregular real behavior. In this way, our work is closer in spirit to recent efforts to capture natural behaviors with simplified empirical models \citep{berman2016predictability, hong2015automated} that are more interpretable than pure black-box formulations.

Moreover, by formulating our model as a generative process, we receive the added benefit that training produces an artificial agent that replicates the tendencies and biases of real players, a key advantage for studies of interactive play. By adding player identity as the input to such a model, we should be able to successfully capture and reproduce a variety of play styles. Finally, we can consider replacing our mixture model by a more structured temporal model like a hierarchical hidden Markov model, in which case our player can simply be viewed as a sequence of discrete (hierarchial) strategic choices with a complex observation model parameterized by a neural network. Our mixture model would then be equivalent to integrating out this HMM layer.

\section{Conclusion}
\label{conclusion_section}
We have proposed a model of inverse reinforcement learning based on inferring latent trajectories for goal states through time. These goal states give rise to observed states via a control model, and their evolution is governed by a Gaussian process determined by a dynamic value function. The model combines two generative approaches: a variational autoencoder (for inferring posteriors over goals given observations) and a gaussian mixture model (for approximating $V$).

Our model provides a structured method for understanding behavior in a dynamic, continuous-control task and allows for behavioral task complexity to scale up with the complexity of neural data. We have shown that such a model is able to reproduce the noisy and heterogenous behavior of real agents engaged in a competitive video game, and that by modeling the value function explicitly, we are able to easily visualize the multimodal distribution over future strategies for any instantaneous configuration of the system. This constitutes a significant improvement over conventional models in psychology, animal behavior, and related fields that seek to model the decisions of agents in dynamic environments, as it dramatically increases model flexibility while retaining the interpretability of a cost function over latent goals. For our example data, the inferred goal states, as well as their associated value functions, give important clues as to the underlying strategies employed by players, as well as offering potential neural correlates of the decision process.

\subsubsection*{Acknowledgments}
We would like to thank Michael Platt for sharing the data used in this work and Caroline Drucker and David Carlson for helpful discussions. This work was funded by NIH grants R01-MH109728 (PI: Platt; JP Co-Investigator) and a BD2K career development award (K01-ES025442) to JP.
\newpage
\bibliography{iqbal_pearson}

\begin{thebibliography}{28}
\providecommand{\natexlab}[1]{#1}
\providecommand{\url}[1]{\texttt{#1}}
\expandafter\ifx\csname urlstyle\endcsname\relax
  \providecommand{\doi}[1]{doi: #1}\else
  \providecommand{\doi}{doi: \begingroup \urlstyle{rm}\Url}\fi

\bibitem[Pearson et~al.(2014)Pearson, Watson, and Platt]{pearson2014decision}
John~M Pearson, Karli~K Watson, and Michael~L Platt.
\newblock Decision making: the neuroethological turn.
\newblock \emph{Neuron}, 82\penalty0 (5):\penalty0 950--965, 2014.

\bibitem[Freeman et~al.(2014)Freeman, Vladimirov, Kawashima, Mu, Sofroniew,
  Bennett, Rosen, Yang, Looger, and Ahrens]{freeman2014mapping}
Jeremy Freeman, Nikita Vladimirov, Takashi Kawashima, Yu~Mu, Nicholas~J
  Sofroniew, Davis~V Bennett, Joshua Rosen, Chao-Tsung Yang, Loren~L Looger,
  and Misha~B Ahrens.
\newblock Mapping brain activity at scale with cluster computing.
\newblock \emph{Nature methods}, 11\penalty0 (9):\penalty0 941--950, 2014.

\bibitem[Pnevmatikakis et~al.(2016)Pnevmatikakis, Soudry, Gao, Machado, Merel,
  Pfau, Reardon, Mu, Lacefield, Yang, et~al.]{pnevmatikakis2016simultaneous}
Eftychios~A Pnevmatikakis, Daniel Soudry, Yuanjun Gao, Timothy~A Machado, Josh
  Merel, David Pfau, Thomas Reardon, Yu~Mu, Clay Lacefield, Weijian Yang,
  et~al.
\newblock Simultaneous denoising, deconvolution, and demixing of calcium
  imaging data.
\newblock \emph{Neuron}, 89\penalty0 (2):\penalty0 285--299, 2016.

\bibitem[Pandarinath et~al.(2017)Pandarinath, O'Shea, Collins, Jozefowicz,
  Stavisky, Kao, Trautmann, Kaufman, Ryu, Hochberg,
  et~al.]{pandarinath2017inferring}
Chethan Pandarinath, Daniel~J O'Shea, Jasmine Collins, Rafal Jozefowicz,
  Sergey~D Stavisky, Jonathan~C Kao, Eric~M Trautmann, Matthew~T Kaufman,
  Stephen~I Ryu, Leigh~R Hochberg, et~al.
\newblock Inferring single-trial neural population dynamics using sequential
  auto-encoders.
\newblock \emph{bioRxiv}, page 152884, 2017.

\bibitem[Zheng et~al.(2016)Zheng, Yue, and Hobbs]{zheng2016generating}
Stephan Zheng, Yisong Yue, and Jennifer Hobbs.
\newblock Generating long-term trajectories using deep hierarchical networks.
\newblock In \emph{Advances in Neural Information Processing Systems}, pages
  1543--1551, 2016.

\bibitem[Moussa{\"\i}d et~al.(2009)Moussa{\"\i}d, Helbing, Garnier, Johansson,
  Combe, and Theraulaz]{moussaid2009experimental}
Mehdi Moussa{\"\i}d, Dirk Helbing, Simon Garnier, Anders Johansson, Maud Combe,
  and Guy Theraulaz.
\newblock Experimental study of the behavioural mechanisms underlying
  self-organization in human crowds.
\newblock \emph{Proceedings of the Royal Society of London B: Biological
  Sciences}, 276\penalty0 (1668):\penalty0 2755--2762, 2009.

\bibitem[Ng et~al.(2000)Ng, Russell, et~al.]{ng2000algorithms}
Andrew~Y Ng, Stuart~J Russell, et~al.
\newblock Algorithms for inverse reinforcement learning.
\newblock In \emph{Icml}, pages 663--670, 2000.

\bibitem[Abbeel and Ng(2004)]{abbeel2004apprenticeship}
Pieter Abbeel and Andrew~Y Ng.
\newblock Apprenticeship learning via inverse reinforcement learning.
\newblock In \emph{Proceedings of the twenty-first international conference on
  Machine learning}, page~1. ACM, 2004.

\bibitem[Dvijotham and Todorov(2010)]{Dvijotham2010-lz}
Krishnamurthy Dvijotham and Emanuel Todorov.
\newblock Inverse optimal control with linearly-solvable {MDPs}.
\newblock In \emph{Proceedings of the 27th International Conference on Machine
  Learning ({ICML-10})}, pages 335--342. machinelearning.wustl.edu, 2010.

\bibitem[Park and Sandberg(1991)]{park1991universal}
Jooyoung Park and Irwin~W Sandberg.
\newblock Universal approximation using radial-basis-function networks.
\newblock \emph{Neural computation}, 3\penalty0 (2):\penalty0 246--257, 1991.

\bibitem[Beal(2003)]{beal2003variational}
Matthew~James Beal.
\newblock \emph{Variational algorithms for approximate Bayesian inference}.
\newblock University of London United Kingdom, 2003.

\bibitem[Wainwright et~al.(2008)Wainwright, Jordan,
  et~al.]{wainwright2008graphical}
Martin~J Wainwright, Michael~I Jordan, et~al.
\newblock Graphical models, exponential families, and variational inference.
\newblock \emph{Foundations and Trends{\textregistered} in Machine Learning},
  1\penalty0 (1--2):\penalty0 1--305, 2008.

\bibitem[Ranganath et~al.(2014)Ranganath, Gerrish, and
  Blei]{ranganath2014black}
Rajesh Ranganath, Sean Gerrish, and David~M Blei.
\newblock Black box variational inference.
\newblock In \emph{AISTATS}, pages 814--822, 2014.

\bibitem[Kucukelbir et~al.(2015)Kucukelbir, Ranganath, Gelman, and
  Blei]{kucukelbir2015automatic}
Alp Kucukelbir, Rajesh Ranganath, Andrew Gelman, and David Blei.
\newblock Automatic variational inference in stan.
\newblock In \emph{Advances in neural information processing systems}, pages
  568--576, 2015.

\bibitem[Rezende et~al.(2014)Rezende, Mohamed, and
  Wierstra]{rezende2014stochastic}
Danilo~Jimenez Rezende, Shakir Mohamed, and Daan Wierstra.
\newblock Stochastic backpropagation and approximate inference in deep
  generative models.
\newblock \emph{arXiv preprint arXiv:1401.4082}, 2014.

\bibitem[Kingma and Welling(2013)]{Kingma2013-td}
Diederik~P Kingma and Max Welling.
\newblock {Auto-Encoding} variational bayes.
\newblock 20~December 2013.

\bibitem[Archer et~al.(2015)Archer, Park, Buesing, Cunningham, and
  Paninski]{Archer2015-ec}
Evan Archer, Il~Memming Park, Lars Buesing, John Cunningham, and Liam Paninski.
\newblock Black box variational inference for state space models.
\newblock 23~November 2015.

\bibitem[Gao et~al.(2016)Gao, Archer, Paninski, and Cunningham]{Gao2016-ck}
Yuanjun Gao, Evan Archer, Liam Paninski, and John~P Cunningham.
\newblock Linear dynamical neural population models through nonlinear
  embeddings.
\newblock 26~May 2016.

\bibitem[Kingma and Ba(2014)]{Kingma2014-pz}
Diederik Kingma and Jimmy Ba.
\newblock Adam: A method for stochastic optimization.
\newblock 22~December 2014.

\bibitem[Chung et~al.(2015)Chung, Kastner, Dinh, Goel, Courville, and
  Bengio]{Chung2015-sy}
Junyoung Chung, Kyle Kastner, Laurent Dinh, Kratarth Goel, Aaron Courville, and
  Yoshua Bengio.
\newblock A recurrent latent variable model for sequential data.
\newblock 7~June 2015.

\bibitem[Carpenter(1988)]{carpenter1988movements}
Roger~HS Carpenter.
\newblock \emph{Movements of the Eyes, 2nd Rev}.
\newblock Pion Limited, 1988.

\bibitem[Todorov and Jordan(2002)]{todorov2002optimal}
Emanuel Todorov and Michael~I Jordan.
\newblock Optimal feedback control as a theory of motor coordination.
\newblock \emph{Nature neuroscience}, 5\penalty0 (11):\penalty0 1226--1235,
  2002.

\bibitem[Sussillo and Abbott(2009)]{sussillo2009generating}
David Sussillo and Larry~F Abbott.
\newblock Generating coherent patterns of activity from chaotic neural
  networks.
\newblock \emph{Neuron}, 63\penalty0 (4):\penalty0 544--557, 2009.

\bibitem[Song et~al.(2016)Song, Yang, and Wang]{song2016training}
H~Francis Song, Guangyu~R Yang, and Xiao-Jing Wang.
\newblock Training excitatory-inhibitory recurrent neural networks for
  cognitive tasks: A simple and flexible framework.
\newblock \emph{PLoS computational biology}, 12\penalty0 (2):\penalty0
  e1004792, 2016.

\bibitem[Thalmeier et~al.(2016)Thalmeier, Uhlmann, Kappen, and
  Memmesheimer]{thalmeier2016learning}
Dominik Thalmeier, Marvin Uhlmann, Hilbert~J Kappen, and Raoul-Martin
  Memmesheimer.
\newblock Learning universal computations with spikes.
\newblock \emph{PLoS computational biology}, 12\penalty0 (6):\penalty0
  e1004895, 2016.

\bibitem[Abbott et~al.(2016)Abbott, DePasquale, and
  Memmesheimer]{abbott2016building}
LF~Abbott, Brian DePasquale, and Raoul-Martin Memmesheimer.
\newblock Building functional networks of spiking model neurons.
\newblock \emph{Nature neuroscience}, 19\penalty0 (3):\penalty0 350--355, 2016.

\bibitem[Berman et~al.(2016)Berman, Bialek, and
  Shaevitz]{berman2016predictability}
Gordon~J Berman, William Bialek, and Joshua~W Shaevitz.
\newblock Predictability and hierarchy in drosophila behavior.
\newblock \emph{Proceedings of the National Academy of Sciences}, 113\penalty0
  (42):\penalty0 11943--11948, 2016.

\bibitem[Hong et~al.(2015)Hong, Kennedy, Burgos-Artizzu, Zelikowsky, Navonne,
  Perona, and Anderson]{hong2015automated}
Weizhe Hong, Ann Kennedy, Xavier~P Burgos-Artizzu, Moriel Zelikowsky,
  Santiago~G Navonne, Pietro Perona, and David~J Anderson.
\newblock Automated measurement of mouse social behaviors using depth sensing,
  video tracking, and machine learning.
\newblock \emph{Proceedings of the National Academy of Sciences}, 112\penalty0
  (38):\penalty0 E5351--E5360, 2015.

\end{thebibliography}

\end{document}